# Gravimetric estimation of the Eötvös components

G. Manoussakis[(1)], R. Korakitis[(1)] and P. Milas[(2)]

[(1)]Dionysos Satellite Observatory and [(2)]Higher Geodesy Laboratory, Department of Surveying, National Technical University of Athens, Iroon Polytechniou 9, Zografos 157 80, Greece, tel. +30–2107722693, fax +30–2107722670.

gmanous@survey.ntua.gr

**Abstract**

The elements of the Eötvös matrix are useful for various geodetic applications, such as the interpolation of the elements of the deflection of the vertical, the determination of gravity anomalies and the determination of geoid heights. A torsion balance instrument is customarily used for the determination of the Eötvös components. In this work, we show that it is possible to estimate the Eötvös components at a point on the Earth's physical surface using gravity measurements at three nearby points, comprising a very small network. In the first part, we present the method in detail, while in the second part we demonstrate a numerical example. We conclude that this method is able to estimate the elements of the Eötvös matrix with satisfactory accuracy.

**1 Introduction**

The Eötvös matrix is the second order derivative of the Earth's gravity potential at a point $P$ and is significant to several applications. For example, it plays an important role in the "Geodetic Singularity Problem": if the determinant of the Eötvös matrix at point $P$ is equal to zero, then it is rank deficient and this classifies point $P$ as a singular point. This means that it is not possible to replace (pseudo)differentials of unholonomic coordinate systems, which are related to moving local astronomical frames, with differentials of holonomic coordinate systems (Grafarend, 1971, Livieratos, 1976). Another application of the Eötvös matrix is the determination of the deflection of the vertical at points on the Earth's physical surface (Völgyesi, 1993). The elements of the Eötvös matrix which are involved are $W_{xx}$, $W_{xy}$ and $W_{yy}$. A third application of the Eötvös matrix is the determination of the geoid undulation by an alternative solution for the astrogeodetic leveling (Völgyesi, 2001), and to determine the gravity anomaly with the help of the elements $W_{xz}$ and $W_{yz}$ (Völgyesi et al., 2005).

The elements of the Eötvös matrix (except $W_{zz}$) are customarily determined by torsion balance measurements at point $P$. The appreciation of the Eötvös matrix is lately increased, since a large number of torsion balance measurements are carried out around the world (Völgyesi, 2015), in order to detect lateral underground mass inhomogeneities and geological fault structures.

The aim of the present work is to develop a method for the estimation of the Eötvös components using a gravimeter instead of a torsion balance instrument. The components will be estimated at a chosen point $S$ on the Earth's physical surface, using gravity measurements at S and three nearby points, comprising a very small network. The proposed method will be described in detail in the next sections.

## 2 Methodology

### 2.1 Estimation of the values of the Eötvös components except $W_{xy}$

Let S be a point on the Earth's physical surface with known geodetic coordinates, gravity value and geometric height. The Earth's gravity potential is expressed in a local Cartesian system $(x, y, z)$, which is centered at point S (point of interest), the $z$ – axis is perpendicular to the equipotential surface passing through point S pointing outwards, the $x$ – axis is tangent to the equipotential surface passing through point S pointing North and the $y$ – axis is tangent to the aforementioned equipotential surface pointing East.

In addition, let A, B, C be three points in the neighborhood of point S (within a few meters) with known local Cartesian coordinates and gravity values. Point A is taken on the $x$ – axis ($y = 0$), point B on the $y$ – axis ($x = 0$) and point C on the $z$ – axis ($x = y = 0$), as in Figure 1.

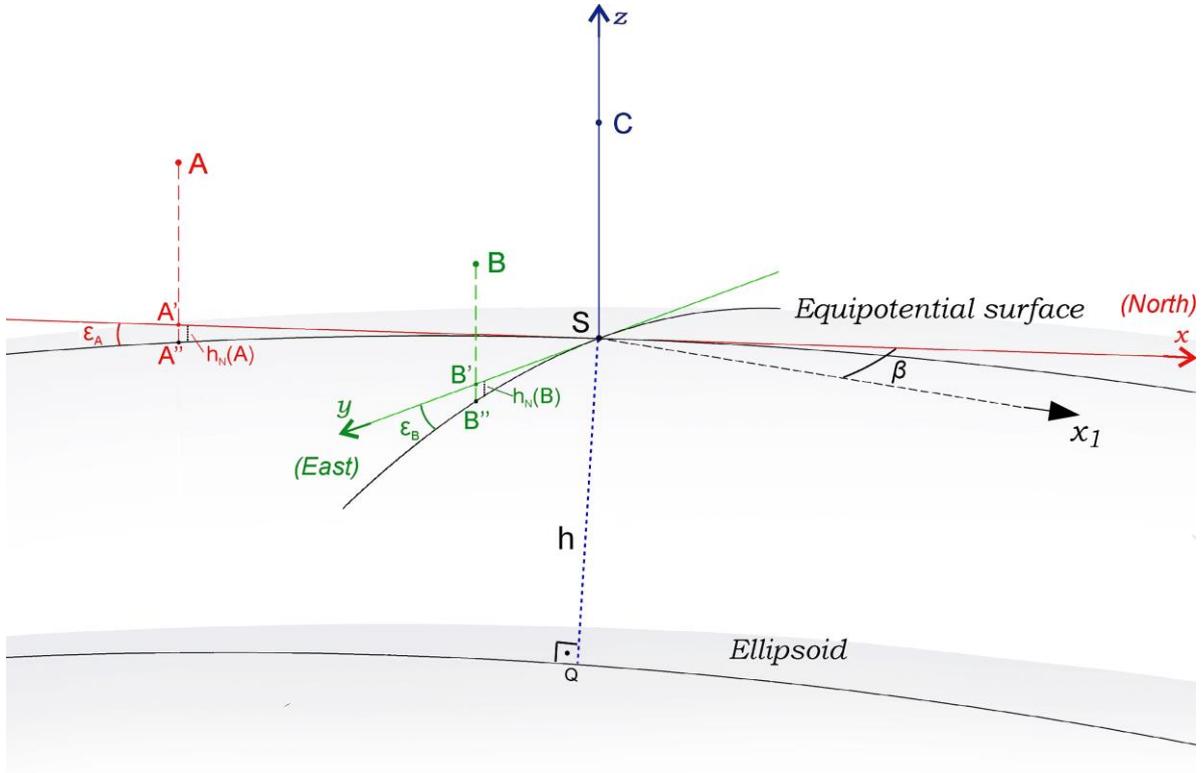

Figure 1

The value of $W_{zz}$ can be directly computed from the gravity measurements at points S and C (see (2.4)). For the other four Eötvös components at point S we proceed as follows:

A parametric vector equation for the actual equipotential surface of point S around this point, expressed in the local Cartesian system, is

$$\bar{s}: \mathfrak{R}^2 \supset D \to \mathfrak{R}^3 : (x, y) \to \bar{s}(x, y) = (x, y, z(x, y)) \qquad (2.1)$$

and the tangent vectors of the equipotential surface are

$$\frac{\partial \bar{s}}{\partial x} \equiv \bar{s}_x = \left(1, 0, -\frac{W_x}{W_z}\right) \tag{2.2}$$

$$\frac{\partial \bar{s}}{\partial y} \equiv \bar{s}_y = \left(0, 1, -\frac{W_y}{W_z}\right) \tag{2.3}$$

The value of $W_{zz}$ at point S is obtained by:

$$W_{zz}(S) \equiv W_{zz} = -\frac{g_C - g_S}{z_C} \tag{2.4}$$

Approximate, temporary values of $W_{xx}$ and $W_{yy}$ at point S can be obtained as follows: Let $x_1, y_1$ be the axes of the principal directions at point S. We set

$$W_{yy}{}^t(S) \equiv W_{yy}{}^t \approx W_{y_1 y_1}(S) \equiv W_{y_1 y_1} := -g_S k_2{}^e(S) \tag{2.5}$$

The quantity $k_2{}^e$ is the principal curvature of the normal equipotential surface in the East-West direction at point S, which is equal to (Manoussakis, 2013):

$$k_2{}^e(S) \equiv k_2{}^e = -\frac{U_{yy}(S)}{\gamma_S} \tag{2.6}$$

where

$$U_{yy}(S) = -\gamma_Q k_2(Q) + \gamma_Q k_2{}^2(Q) h_S \tag{2.7}$$

In (2.7), point Q is the projection of point S on the ellipsoid of revolution along the vertical line to the ellipsoid, $k_2$ is the principal curvature of the ellipsoid along the East – West direction and $U$ stands for the Eötvös matrix of the normal gravity field. From Poisson's equation we can also estimate $W^t_{xx}$ at point S:

$$W_{xx}{}^t = 2\omega^2 - W_{zz} - W_{yy}{}^t \tag{2.8}$$

We now proceed to estimate $W_{xz}$ and $W_{yz}$. Let A´ be the projection of point A on the tangent plane of the equipotential surface at point S and A´´ the intersection of the line AA´ with the actual equipotential surface passing through point S (figure 1). Using a Taylor series expansion, we get

$$\bar{s}(A'') - \bar{s}(A) = \bar{s}_x(S) x_A + \bar{s}_y(S) y_A + \frac{1}{2}(\bar{s}_{xx}(S) x_A{}^2 + 2\bar{s}_{xy}(S) x_A y_A + \bar{s}_{yy}(S) y_A{}^2) \tag{2.9}$$

Multiplying both sides with the unit normal vector and, having in mind that point A is on the $xz$ - plane, we end up with:

$$h_N(A) \equiv |z(A'')| = \left|\frac{1}{2} L_{act} x_A^2\right| \tag{2.10}$$

where $L_{act}$ is an element of the second fundamental form of the actual equipotential surface at point S, i.e.

$$L_{act} = \langle \overline{N}, \overline{s}_{xx} \rangle \tag{2.11}$$

where

$$\overline{N} = \frac{1}{|\overline{s}_x \times \overline{s}_y|} (\overline{s}_x \times \overline{s}_y) \tag{2.12}$$

The value of $L_{act}$ is given by

$$L_{act} = -\frac{W_{xx}^t}{g_S} \tag{2.13}$$

The angle $\varepsilon_A$ (see figure 1) is computed by

$$\tan \varepsilon_A \cong \varepsilon_A = \frac{|z(A'')|}{|x_A|} = \left|\frac{1}{2} L_{act} x_A\right| \tag{2.14}$$

Since A′ is the projection of point A on the tangent plane of the actual equipotential surface at point S ($z_{A'} = 0$), we obtain

$$g_{A'} = g_A - W_{zz}(z_{A'} - z_A) = g_A + W_{zz} z_A \tag{2.15}$$

$$W_z(A') = -g_{A'} \cos \varepsilon_A \tag{2.16}$$

The values of $W_z$ are known at points A′ and S, so the value of the second order partial derivative $W_{xz}$ at point S is given by:

$$W_{xz} = \frac{W_z(A') - W_z(S)}{x_A} = \frac{g_S - g_{A'} \cos \varepsilon_A}{x_A} \tag{2.17}$$

Repeating the above procedure for a point B on the West – East direction, we have similar relations, which lead to the estimation of the value of the second order partial derivative $W_{yz}$ at point S:

$$h_N(B) \equiv |z(B'')| = \left|\frac{1}{2} N_{act} y_B^2\right| \tag{2.18}$$

$$N_{act} = \langle \overline{N}, \overline{s}_{xx} \rangle = -\frac{W_{yy}^t}{g_S} \tag{2.19}$$

$$\tan \varepsilon_B \cong \varepsilon_B = \frac{|z(B'')|}{|y_B|} = \left|\frac{1}{2} N_{act} y_B\right| \tag{2.20}$$

Since B′ is the projection of point B on the tangent plane of the actual equipotential surface at point S, we obtain

$$g_{B'} = g_B + W_{zz}(z_{B'} - z_B) = g_B + W_{zz} z_B \tag{2.21}$$

$$W_z(B') = -g_{B'} \cos\varepsilon_B \tag{2.22}$$

Therefore, the second order partial derivative $W_{yz}$ at point S is given by

$$W_{yz} = \frac{W_z(B') - W_z(S)}{y_B} = \frac{g_S - g_{B'} \cos\varepsilon_B}{y_B} \tag{2.23}$$

## 2.2 Estimation of the value of the Eötvös component $W_{xy}$

Let $\beta$ be the angle of a principal direction $x_1$ with the $x$ – axis (figure 1). The transformation between the two coordinate systems is given by

$$\begin{bmatrix} x \\ y \\ z \end{bmatrix} = \begin{bmatrix} \cos\beta & -\sin\beta & 0 \\ \sin\beta & \cos\beta & 0 \\ 0 & 0 & 1 \end{bmatrix} \begin{bmatrix} x_1 \\ y_1 \\ z_1 \end{bmatrix} \tag{2.24}$$

Using (2.24), we can relate the gradients of the potential between the two systems:

$$W_{x_1} = W_x \frac{\partial x}{\partial x_1} + W_y \frac{\partial y}{\partial x_1} + W_z \frac{\partial z}{\partial x_1} = W_x \cos\beta + W_y \sin\beta \tag{2.25}$$

and

$$W_{xx}{}^t\left(-\frac{1}{2}\cos 2\beta\right) + W_{xy}{}^t \cos 2\beta + W_{yy}{}^t\left(\frac{1}{2}\sin 2\beta\right) = W_{x_1 y_1} \tag{2.26}$$

By definition, the right-hand side of (2.26) is equal to zero. From Euler's theorem, we can determine the value of the sectional curvature $k_n$ of the actual equipotential surface at point S as a function of an angle $a$ ($a = 0$ along the $x_1$ axis), i.e.

$$k_n(S) = k_1(S)\cos^2 a + k_2(S)\sin^2 a \tag{2.27}$$

where $k_1(S)$ and $k_2(S)$ are the principal curvatures at point S. For $a = \pi/4$, the sectional curvature at point S is equal to the mean curvature of the equipotential surface at the same point. The mean curvature is an invariant of the equipotential surface and in both Cartesian systems its expression stays the same, i.e.

$$H(S) = -\frac{W_{x_1 x_1} + W_{y_1 y_1}}{2g_S} = -\frac{W_{xx}^{\ t} + W_{yy}^{\ t}}{2g_S} \tag{2.28}$$

Now let E be a fictitious point, close to point S, such that the angle between the $x_1$ axis and the line segment SA is equal to $\pi/4$. In addition, let $x_E$ and $y_E$ be the coordinates of the point E. Then it holds that

$$k_n(S) = \frac{L_{act} x_E^2 + 2M_{act} x_E y_E + N_{act} y_E^2}{E_{act} x_E^2 + G_{act} y_E^2} \tag{2.29}$$

The quantities $L_{act}$ and $N_{act}$ are given from (2.13) and (2.19), respectively. From (2.2) and (2.3) we have that

$$E_{act} = \left\langle \left(1, 0, -\frac{W_x}{W_z}\right)_S, \left(1, 0, -\frac{W_x}{W_z}\right)_S \right\rangle = 1 \tag{2.30}$$

$$G_{act} = \left\langle \left(0, 1, -\frac{W_y}{W_z}\right)_S, \left(0, 1, -\frac{W_y}{W_z}\right)_S \right\rangle = 1 \tag{2.31}$$

$M_{act}$ is defined as

$$M_{act} = \langle \overline{N}, \overline{s}_{xy} \rangle \tag{2.32}$$

hence

$$M_{act} = -\frac{W_{xy}}{g_S} \tag{2.33}$$

Substituting (2.13), (2.19), (2.28), (2.30), (2.31) and (2.33) into (2.29), we get

$$\frac{W_{xx}^{\ t} + W_{yy}^{\ t}}{2} = \frac{W_{xx}^{\ t} x_E^2 + 2 W_{xy}^{\ t} x_E y_E + W_{yy}^{\ t} y_E^2}{x_E^2 + y_E^2} \tag{2.34}$$

All partial derivatives are referred to point S. We can assume that $y_E \neq 0$, so dividing both nominator and denominator by $y_E$ we get a different form of the above equation:

$$\frac{W_{xx}^{\ t} + W_{yy}^{\ t}}{2} = \frac{W_{xx}^{\ t}\left(\frac{x_E}{y_E}\right)^2 + 2 W_{xy}^{\ t} \frac{x_E}{y_E} + W_{yy}^{\ t}}{\left(\frac{x_E}{y_E}\right)^2 + 1} \tag{2.35}$$

Let $\theta$ be the angle which is defined by

$$\tan\theta = \frac{x_E}{y_E} \tag{2.36}$$

Then

$$\theta = \frac{\pi}{2} - \left(\frac{\pi}{4} + \beta\right) = \frac{\pi}{4} - \beta \tag{2.37}$$

Setting

$$t = \tan\left(\frac{\pi}{4} - \beta\right) \tag{2.38}$$

equation (2.35) becomes

$$\frac{W_{xx}{}^t + W_{yy}{}^t}{2} = \frac{W_{xx}{}^t t^2 + 2W_{xy}{}^t t + W_{yy}{}^t}{t^2 + 1} \tag{2.39}$$

Equations (2.26) and (2.39) comprise a system of two equations with two unknowns, the angle $\beta$ and the temporary value of $W_{xy}$. In order to solve this system, we isolate the term $4W_{xy}t$ from (2.39) and we obtain:

$$4W_{xy}{}^t t = (W_{yy}{}^t - W_{xx}{}^t)t^2 - (W_{yy}{}^t - W_{xx}{}^t) \tag{2.40}$$

From elementary trigonometry and using (2.38) we get

$$\cot 2\beta = \tan\left(\frac{\pi}{2} - 2\beta\right) = \frac{2t}{1 - t^2} \tag{2.41}$$

Dividing (2.26) by sin2β (thus excluding the values 0 and π/2 for β) we have

$$W_{xx}{}^t \frac{2t}{1 - t^2} - 2W_{xy}{}^t \frac{2t}{1 - t^2} + W_{yy}{}^t = 0 \tag{2.42}$$

After some manipulations, we end up with the following relation

$$2W_{xx}{}^t t - 4W_{xy}{}^t t + W_{yy}{}^t - W_{yy}{}^t t^2 = 0 \tag{2.43}$$

Combining (2.40) with (2.43), the following algebraic equation is formed

$$(W_{xx}{}^t - 2W_{yy}{}^t)t^2 + 2W_{xx}{}^t t - (W_{xx}{}^t - 2W_{yy}{}^t) = 0 \tag{2.44}$$

From (2.44) we can obtain the two real values of $t$ :

$$t_{1,2} = \frac{-W_{xx}{}^t \pm \sqrt{(W_{xx}{}^t)^2 + (W_{xx}{}^t - 2W_{yy}{}^t)^2}}{W_{xx}{}^t - 2W_{yy}{}^t} \tag{2.45}$$

Following (2.38), we can now obtain the values of the angle $\beta$ of the principal directions

$$\beta_1 = \frac{\pi}{4} - \arctan\left[\frac{-W_{xx}{}^t + \sqrt{(W_{xx}{}^t)^2 + (W_{xx}{}^t - 2W_{yy}{}^t)^2}}{W_{xx}{}^t - 2W_{yy}{}^t}\right] \qquad (2.46)$$

and

$$\beta_2 = \frac{\pi}{4} - \arctan\left[\frac{-W_{xx}{}^t - \sqrt{(W_{xx}{}^t)^2 + (W_{xx}{}^t - 2W_{yy}{}^t)^2}}{W_{xx}{}^t - 2W_{yy}{}^t}\right] \qquad (2.47)$$

We should note that the angle $\beta$ corresponds to the angle between two lines and not between two axes. We need additional independent information about the special transformation $x = x(x_1, y_1)$, $y = y(x_1, y_1)$ which holds around point S in order to find the principal axes $x_1$ and $y_1$. Therefore, we cannot distinguish which of the two above values refers to the $x_1$ axis. In what follows, we chose the smaller of the two values of $\beta$, which leads us to the final value of the parameter $t$ (see eq.(2.38)).

Finally, using (2.40) we are now able to estimate a temporary value of $W_{xy}$:

$$W_{xy}{}^t = \frac{(W_{yy}{}^t - W_{xx}{}^t)(t^2 - 1)}{4t} \qquad (2.48)$$

## 2.3 Refinement of the values of the Eötvös components $W_{xx}$, $W_{yy}$ and $W_{xy}$

From the transformation described in (2.24) we can construct relations similar to (2.26) for the other surface derivatives, so we end up with the following system of equations:

$$W_{xx}\cos^2\beta + W_{xy}\sin 2\beta + W_{yy}\sin^2\beta = W_{x_1 x_1} \qquad (2.49)$$

and

$$W_{xx}\sin^2\beta + W_{xy}(-\sin 2\beta) + W_{yy}\cos^2\beta = W_{y_1 y_1} \qquad (2.50)$$

Recalling our assumption for the estimate of $W_{yy}$ (see (2.5)), we introduce the temporary values $W^t{}_{xx}$ and $W^t{}_{yy}$ to the right-hand side of (2.49) and (2.50), i.e.

$$\begin{aligned} W_{x_1 x_1} &= W_{xx}{}^t \\ W_{y_1 y_1} &= W_{yy}{}^t \end{aligned} \qquad (2.51)$$

Thus, we now have a system of two equations, which has as unknowns the refined values of $W_{xx}$ and $W_{yy}$. Its solution is:

$$W_{xx} = \frac{W_{xx}{}^t \cos^2 \beta - W_{xy}{}^t \sin 2\beta - W_{yy}{}^t \sin^2 \beta}{\cos 2\beta} \qquad (2.52)$$

and

$$W_{yy} = \frac{W_{yy}{}^t \cos^2 \beta + W_{xy}{}^t \sin 2\beta - W_{xx}{}^t \sin^2 \beta}{\cos 2\beta} \qquad (2.53)$$

A further refinement of the values of $W_{xx}$, $W_{yy}$ and $W_{xy}$ can be done by repeating the calculations from (2.45) up to (2.53), using the new values of the Eötvös components.

## 3 Numerical Test

In order to examine the performance of the proposed method, we made a numerical simulation. We chose 12 arbitrary points on the Earth's physical surface, as in Table 1, and we computed the simulated Eötvös matrix components and gravity values at those (S) and at nearby points (A, B, C), using the EGM96 gravity model. Then, we applied the proposed method to estimate the Eötvös components from the gravity values.

| $\varphi^0$ (geodetic latitude, $^0$) | $\lambda^0$ (geodetic longitude, $^0$) | $h$ (geometric height, $m$) |
|---|---|---|
| 57.20 | -2.30 | 140.00 |
| 50.40 | -4.10 | 70.00 |
| 43.30 | -0.40 | 260.00 |
| 37.30 | -6.00 | 40.00 |
| 52.20 | 4.50 | 0.00 |
| 57.00 | 7.90 | 50.00 |
| 49.90 | 11.60 | 390.00 |
| 37.97 | 23.78 | 225.00 |
| 47.80 | 21.70 | 120.00 |
| 55.20 | 30.20 | 170.00 |
| 45.00 | 39.00 | 20.00 |
| 33.30 | 44.50 | 40.00 |

Table 1

The x, y coordinates of the nearby points were randomly assigned values in the range [-5m , 5m], while the z coordinates had values in the range [0.5m , 2m]. For the computation of the Eötvös matrix components, the model gravity values at all points were accurate to 1 nanogal. Using these values, in Table 2 below we summarize the maximum differences found between the simulated (from EGM96) and the estimated Eötvös matrix components, in Eötvös units.

| Components | x | y | z |
|---|---|---|---|
| x | -0.60 up to 5.54 | -3.30 up to 4.13 | -0.002 up to 0.003 |
| y |  | -5.54 up to 0.60 | -0.002 up to 0.003 |
| z |  |  | -0.006 up to 0.004 |

Table 2

Using exactly the same points but rounded-off values of the gravity measurements to the nearest µgal, which today is a realistic accuracy level, we obtained the maximum differences shown in Table 3.

| Components | x | y | z |
|---|---|---|---|
| x | -1.11  up to  6.64 | -3.89  up to  4.23 | -1.18  up to  2.35 |
| y |  | -4.77  up to  1.01 | -1.42  up to  1.78 |
| z |  |  | -2.64  up to  1.58 |

Table 3

## 4 Discussion and conclusions

In this work, we outlined a method for the estimation of the components of the Eötvös matrix using local gravity measurements. We chose a point on the Earth's physical surface (point of interest), with known geometric height, and three neighboring points and obtained gravity measurements at all points. We estimated the value of $W_{zz}$ from the gravity measurements and initial values of the surface derivatives $W_{xx}$ and $W_{yy}$, using an approximation for the curvature of the actual equipotential surface along the East – West direction. The values of $W_{xx}$ and $W_{yy}$ were then used for the estimation of $W_{xz}$ and $W_{yz}$. They were also used to compute an approximate value of $W_{xy}$, using Euler's theorem and the invariance of the mean curvature of the equipotential surface. Finally, the value of $W_{xy}$ was used to compute refined values of $W_{xx}$, $W_{yy}$ and $W_{xy}$.

The proposed method was tested by a numerical example, choosing various points on the Earth's physical surface, scattered on a very wide area, and simulating the gravity measurements and the Eötvös components using the EGM96 gravity model. We determined the maximum differences found between the computed and the estimated components of the Eötvös matrix. Despite the approximation made for the curvature of the actual equipotential surface, we have shown that the gravimetric estimation of the components of the Eötvös matrix is quite satisfactory. Using gravity values at the accuracy level of current gravimeters, we estimated $W_{xy}$ and the vertical gradient components of the Eötvös matrix with an accuracy comparable to the one obtained by torsion balance instruments. It is to be noted that, in principle, the values of the vertical gradient components $W_{xz}$, $W_{yz}$ and $W_{zz}$ were very accurate, while the accuracy of the values of $W_{xx}$, $W_{yy}$ and $W_{xy}$ may be further improved, through a better estimation of the curvature.

## References

Grafarend E. 1971: The object of anholonomity and a generalized Riemannian geometry for Geodesy, Boll. Geof. Teor. Applic., 13, 241 – 253.

Hofmann – Wellenhoff B., Moritz H., 2006: Physical Geodesy, Second Edition, Springer Wien – New York, 82, 233.

Livieratos E. 1976: On the geodetic singularity problem, Manuscripta Geodetica, 1, 269 – 292.

Manoussakis G., 2013: Estimation of the normal Eötvös matrix for low geometric heights. Acta Geodetica et Geophysica, 48, 2, 179–189.

Toth Gy., Rozsa Sz., Adam J. and Tziavos I.N., 2001: Gravity Field Recovery from Torsion Balance Data Using Collocation and Spectral Methods, Presented at the EGS XXVI General Assembly, Nice, France, March 26 - 30.

Völgyesi L. 1993: Interpolation of the deflection of the vertical based on gravity gradients, Periodica Polytechnica Ser. Civil Eng., 37, 2, 137 – 166.

Völgyesi L. 1998: Geoid Computations Based on Torsion Balance Measurements, Reports of the Finnish Geodetic Institute 98, 4, 145 – 151.

Völgyesi L. 2001: Local geoid determination based on gravity gradients, Acta Geodetica et Geophysica Hungarica, 36, 2, 153 – 162.

Völgyesi L, Toth Gy., Csapo G. 2005: Determination of gravity anomalies from torsion balance measurements, Gravity Geoid and Space Missions, volume 129 of the IAG Symposia, 292 – 297.

Völgyesi L. 2015: Renaissance of Torsion Balance Measurements in Hungary, Periodica Polytechnica Ser. Civil Engineering, 59 (4), 459 – 464.